\documentclass[%
 reprint,
superscriptaddress,
nofootinbib,
nobibnotes,
 amsmath,amssymb,
 aps, prl, twocolumn
]{revtex4-1}
\usepackage[colorlinks=true,citecolor=blue,linkcolor=blue,breaklinks=true]{hyperref}
\usepackage{graphicx} 
\usepackage{color}
\usepackage{orcidlink}
\usepackage{amsmath}
\usepackage{slashed}
\usepackage{xspace}
\usepackage{multirow}

\date{\today}

\newcommand{\Zbkg}{{\ensuremath{Z}{\it-bkg}}\xspace}
\newcommand{\tbkg}{\ensuremath{t}{\it-bkg}\xspace}

\begin{abstract}

Following the HL-LHC era, proposed lepton colliders highlight the need to study various important Higgs boson production mechanisms to precisely probe the Standard Model Higgs sector. We propose a novel mechanism  $e^\pm \gamma \rightarrow  {\bar \nu}_e (\nu_e) H W^\pm$, which can be useful to study Higgs boson properties.
This channel is relatively free from the background and
can be used to measure the Higgs boson properties, in particular $WWH$ coupling. We examine the viability of this production mechanism. We show that the process can be observed at the planned FCC-ee with the center-of-mass energy of $365$ GeV. At the center-of-mass energy of $500$ GeV, the process can be observed within a few months of the operation.
We use an in-house Monte Carlo event generator that simultaneously incorporates the photon distribution and electron/positron distribution. Our work is also a step towards realistic simulations of lepton- and photon-initiated processes at lepton colliders.

\end{abstract}

\begin{document}
\preprint{IMSc/2026/01}
\title{Associated Higgs production in lepton-photon collisions at FCC-ee}

\author{Pankaj Agrawal\,\orcidlink{0000-0003-3148-3087}}
\email{pankaj.agrawal@tcgcrest.org}
\affiliation{Center for Quantum Engineering, Research, and Education, TCG CREST, Bidhan Nagar, Kolkata - 700091, India}

\author{Biswajit Das\,\orcidlink{0000-0003-1448-2250}}
\email{biswajitd@imsc.res.in}

\author{Tousik Samui\,\orcidlink{0000-0002-1485-6155}}
\email{tousiks@imsc.res.in}
\affiliation{The Institute of Mathematical Sciences, IV Cross Road, CIT Campus, Taramani, Chennai 600113, India.}

\maketitle

The modern understanding of the building blocks of nature and their interactions is successfully described by the Standard Model (SM), which is
based on the gauge group $SU(3)_C \otimes SU(2)_L \otimes U(1)_Y$. Among the Standard Model particles, the Higgs boson plays a central role as
it explains the mass generation of elementary particles via Electroweak symmetry breaking.
The SM Higgs sector parameters are not fully determined experimentally\,\cite{ATLAS:2024fkg,CMS:2026nce}. Therefore, it is important to explore the Higgs sector with new production channels\,\cite{deBlas:2019rxi}.
The SM has been widely tested in the last five decades, and no significant deviation from its predictions has been reported so far. Despite its enormous success in its internal consistency, the SM as a standalone theory has limitations in explaining certain phenomena of nature, 
{\it e.g.}, neutrino masses and oscillations~\cite{Gonzalez-Garcia:2002bkq,Gonzalez-Garcia:2015qrr}, the existence of dark matter and dark energy~\cite{Jee_2007,1970ApJ...159..379R,LZ:2022lsv,Billard:2021uyg,SupernovaCosmologyProject:1998vns,SupernovaSearchTeam:1998fmf}, the absence of CP violation in the strong sector~\cite{Peccei:1977hh}, and the baryon asymmetry~\cite{Gavela:1994dt,Huet:1994jb} of the universe. One thus requires new physics beyond the SM to seek answers to these questions.

High-energy colliders provide a perfect playground for exploring such new physics scenarios, as they have played an instrumental role in establishing the particle content and fundamental structure of the SM. 
Their contribution includes the discovery of particles, establishing interactions, and the precise measurement of SM parameters.
Given the tremendous success of collider experiments, including the currently running LHC, in advancing our knowledge of particle physics, we step into a new era of future colliders. Several new collider facilities have been proposed to probe the SM and search for physics beyond it with unprecedented precision. 
Proposed high-energy lepton colliders such as the International Linear Collider~\cite{ILC:2019gyn}, the Compact Linear Collider~\cite{Wilson:2004qr}, and the Circular Electron-Positron Collider~\cite{Yang:2025wew, CEPCStudyGroup:2025kmw} aim to perform high-precision measurements of the Higgs boson and electroweak observables. In addition, the concept of a muon collider~\cite{IMCC,Delahaye:2019omf,Accettura:2023ked,Ghosh:2023xbj,Skoufaris:2024okx,InternationalMuonCollider:2024jyv,InternationalMuonCollider:2025sys,Ghosh:2025gdx,De:2025dpo} has recently gained significant attention due to its potential to reach very high center-of-mass energies while maintaining the advantages of a lepton collider.

Among these future facilities, the Future Circular Collider (FCC) program \cite{FCC:2018byv, FCC:2025uan, FCC:2025lpp} proposed at CERN represents one of the most ambitious next-generation collider projects. The FCC program is proposed to proceed in stages, beginning with an electron-positron collider (FCC-ee) operating at center-of-mass energies ranging from the
$Z$-pole to the top-quark pair production threshold\,\cite{FCC:2025uan}, followed by a hadron collider (FCC-hh) operating at energies of order 100 TeV. The FCC-ee is expected to provide extremely clean experimental conditions with very low QCD backgrounds, making it an ideal environment for precision studies of the Higgs boson, electroweak interactions, and rare processes.

In view of these future facilities, it is important to explore as many physics scenarios as possible to fully exploit their potential. 
In particular, the Higgs-sector couplings -- $HHH$, $HHHH$, $VVH$, $VVHH$ ($V\equiv W,Z$)--need to be probed more precisely, with ongoing efforts to constrain them through standard processes at the LHC and FCC-ee~\cite{Cepeda:2019klc,qtz8-bkd4, wjcy-1qk6,Agrawal:2019bpm,PhysRevD.97.036006,AGRAWAL2021136461,Agrawal:2025aie}.
Beyond the standard benchmark processes, novel production mechanisms and signatures should be identified in advance so that detector design, trigger strategies, and analysis frameworks can be optimized accordingly. This is particularly crucial since modifications to large experimental setups become challenging once the design is finalized. It is therefore timely to investigate new channels that can be probed at the FCC-ee.

One such interesting direction arises from lepton-photon-initiated processes. At FCC-ee energies, photons radiated (in parallel) from the incoming leptons via initial state radiation (ISR) can carry substantial energy and participate in hard scattering processes. In this regime, the photon can effectively be treated as a partonic constituent of the lepton, allowing for the consideration of an effective photon distribution function. As a result, lepton-photon initiated processes can give non-negligible contributions at high energies.

\begin{figure}[h]
\begin{center}
\includegraphics[width=\columnwidth]{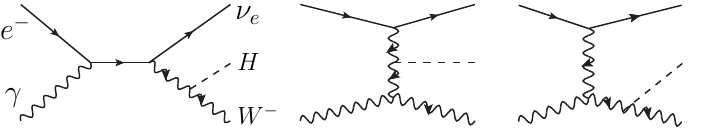}
\end{center}
\caption{Feynman diagrams for the process $e^-\gamma\rightarrow \nu_e H W^-$.}
\label{fig:feyn}
\end{figure}

In this Letter, we focus on the processes
\begin{equation}
	e^- \gamma \rightarrow  \nu_e H W^-, {\rm\qquad and\quad} \gamma e^+ \rightarrow \bar\nu_e H W^+.
\end{equation}
A set of representative Feynman diagrams for these processes is given in Fig.~\ref{fig:feyn}. The Higgs boson predominantly decays into a pair of $b$-jets. We consider the signature of this process when the $W$-boson decays into an electron/muon and associated neutrinos. With these decay modes, the signature of this process is `isolated electron/muon and two $b$-jets'. There will also be missing transverse energy ($\slashed{E}_T$).
The purpose of the study of this channel is two-fold. Firstly, it is a novel channel for the Higgs boson production, and it has not been explored in any lepton collider. This is only possible due to the high energy available at the FCC-ee. Secondly, once this is discovered, this process will play a role as a background for the processes involving a single lepton in the search for new physics beyond the SM. In this Letter, we will focus on the former aspect, as it is a novel channel and an additional channel for the Higgs boson at the FCC-ee. This process is relatively background-free, allowing one to study the properties of the Higgs boson.
This process features the $WWH$ coupling, and the clean signal makes it well-suited for probing this interaction. This coupling has been explored at the LHC \cite{ATLAS:2024fkg,CMS:2026nce}. The main processes contributing to the measurements are -- the VBF and associated production mechanisms; the decay of the Higgs boson to $WW^{*}$; at the one-loop level $H \to \gamma \gamma, \gamma Z$. The uncertainty in the measurement is about $5-6\%$ \cite{ATLAS:2024fkg}.
At the HL-LHC, one can measure this coupling with about $1-2\%$ of precision \cite{Cepeda:2019klc}. A key advantage of the LHC is its high event rate; however, this comes with considerable background contamination. At the FCC-ee, this precision can be improved to $0.5\%$ of precision \cite{FCC:2025lpp}. This bound is mainly due to the VBF fusion production mechanism. Higgs production via $W$-boson fusion is contaminated by $ZH$ production with $Z\rightarrow \nu\bar{\nu}$, making it difficult to isolate the $WWH$ coupling, whereas the proposed process allows a clean and precise probe of the $WWH$ coupling due to negligible background.

There are two major backgrounds which lead to a single charged lepton and two $b$-jets along with non-negligible $\slashed{E}_T$, and
they arise from the processes
\begin{align}
              & e^- \gamma \rightarrow \nu_e Z W^-, & \gamma e^+ \rightarrow \bar\nu_e Z W^+,  \tag{\Zbkg} \\ 
{\rm and\ \ } & e^- \gamma \rightarrow \nu_e \bar{t} b, &\quad\gamma e^+ \rightarrow \bar\nu_e t \bar{b}. \qquad \tag{\tbkg} 
\end{align}
In the \tbkg\ case, the decay $t (\bar{t}) \rightarrow W^+ b (W^-\bar b)$, and in the \Zbkg\ case, the decay $Z\rightarrow b\bar b$, yield an event topology identical to the signal. Additionally, $Z\to c\bar c$ decay in \Zbkg\ case also contributes as a background due to the non-negligible $c \rightarrow b$ misidentification. Henceforth, any reference to \Zbkg includes both the $Z \to b\bar{b}$ and $Z \to c\bar{c}$ modes combined.

As explained previously, the framework we adopt is based on the treatment of the parton distribution functions (PDFs)\,\cite{Frixione:2021zdp}, where a photon or a lepton plays the role of parton and carries a fraction of the initial particle's momentum to participate in the hard scattering. 
Following Ref.~\cite{Skrzypek:1990qs,Cacciari:1992pz,Skrzypek:1992vk,Frixione:2021zdp}, we implement PDFs for leptons and photons relevant to the FCC-ee environment. The photon PDF associated with QED emission is computed using the Improved Weizs\"acker–Williams (IWW) approximation\,\cite{Frixione:1993yw}. For the electron or positron, PDFs are incorporated considering pure QED splitting.
One can also employ PDFs by incorporating the full SM splitting, as implemented in LePDF~\cite{Garosi:2023bvq}. Throughout this study, we work with pure QED PDF sets at LO+LL accuracy\,\cite{Skrzypek:1990qs,Cacciari:1992pz,Skrzypek:1992vk}.

A crucial component of the computation is the convolution of the partonic cross section with the electron or positron PDFs during event generation. With the PDFs, the cross-section for a hard collision of $e^-$ and $e^+$ beams carrying momenta $p_{e^-}$ and $p_{e^+}$, respectively, can be written as
\begin{align}
	d\sigma(p_{e^-},p_{e^+})\!=\!\!\sum_{ij}\!\int\!&dx_- dx_+ \Gamma_{i/{e^-}}(x_-,\mu^2) \Gamma_{j/{e^+}}(x_+,\mu^2) \nonumber\\ 
     &\times\  d\hat{\sigma}_{ij}(x_-p_{e^-},x_+p_{e^+},\mu^2).
\end{align}
In this equation, the $\Gamma_{i/{e^\pm}}$ is the PDF of the {\it parton} $i$ from $e^\pm$ and the $x_\pm$ are the fraction of momenta carried by the parton and $\hat{\sigma}_{ij}$ is the partonic cross-section.
The PDFs, $\Gamma_{e^\pm/e^\pm}(x)$, exhibit a steep rise as the momentum fraction approaches $x \sim 1$, and in fact have singular behavior at $x = 1$. The singular structure of the lepton PDF behaves as $\sim \frac{1}{(1-x)^\beta}$, where $\beta$ for QED emission is given by $\beta = 1 - \frac{\alpha}{\pi}\left(\log\frac{\mu^2}{m_e^2} - 1\right)$ at LO+LL accuracy for a given scale $\mu$. A typical value is $\beta \simeq 0.95$ for the energy range considered in this work. This behavior poses significant challenges for stable numerical evaluation using Monte Carlo techniques.
Since the dominant contribution arises from the region $x \sim 1$, direct numerical integration over $x$ leads to sizable uncertainties, even when using quadruple-precision arithmetic. Achieving a target numerical accuracy below $1\%$ would require an extremely high level of arithmetic precision $\sim\mathcal{O}(10^{-50})$, which is computationally impractical.

To address this issue, we follow the approach of Ref.~\cite{Frixione:2021zdp}, introducing a variable transformation $y = \left(\frac{1 - x}{1 - x_0}\right)^{\beta}$ together with a redefinition of the PDFs, $\Gamma(x) = \frac{\bar{\Gamma}(x)}{(1 - x)^{\beta}}$. We first compute $\bar{\Gamma}(x)$ by multiplying the original PDF by $(1 - x)^{\beta}$. This transformation effectively regularizes the divergence, rendering the PDF behavior near $x \sim 1$ significantly flatter and improving numerical stability.
In contrast, the photon PDF is well behaved near $x \sim 1$ and vanishes at $x = 1$. Thus, no transformation is required for the photon PDF. Both  PDFs also exhibit a divergence at $x \sim 0$; this region is usually excluded by imposing a minimum invariant mass or minimum transverse momentum cut on the final state.
This combined treatment successfully mitigates numerical instabilities arising from the singular behavior of lepton PDFs near $x \sim 1$.

We further note that, for precise predictions at the FCC-ee, it is also important to account for beam-related effects such as \textit{bremsstrahlung}\,\cite{Frixione:2021zdp}. The inclusion of these effects is beyond the scope of the present work; however, we plan to investigate such effects in future studies.

We perform the calculations and parton-level event generation using an in-house framework. In particular, Feynman diagrams required for the computation are generated using {\tt FeynArts}~\cite{Hahn:2000kx}, while the matrix elements are computed with an in-house helicity amplitude generator based on {\tt FORM}\cite{Ruijl:2017dtg}. The phase-space
integration is performed within a Monte Carlo simulation using {\tt AMCI}\,\cite{VESELI19989} with
dedicated phase-space routines for the signal and each background process. 
Finally, the generated event samples are stored in Les Houches Event (LHE) format\cite{Alwall:2006yp} implemented within the in-house routine. 

The masses of the heavy particles used in this study are $m_W=80.419$ GeV, $m_Z=91.188$ GeV, $m_t=173.0$ GeV, and $m_H=125.0$ GeV and the branching ratio of $H\to b\bar b$ is taken to be 58\%.
We have considered one-loop running of the QED coupling $\alpha$, and its value at $m_H$ is $1/\alpha=128.829$.

\begin{figure}
\begin{center}
\includegraphics[width=\columnwidth]{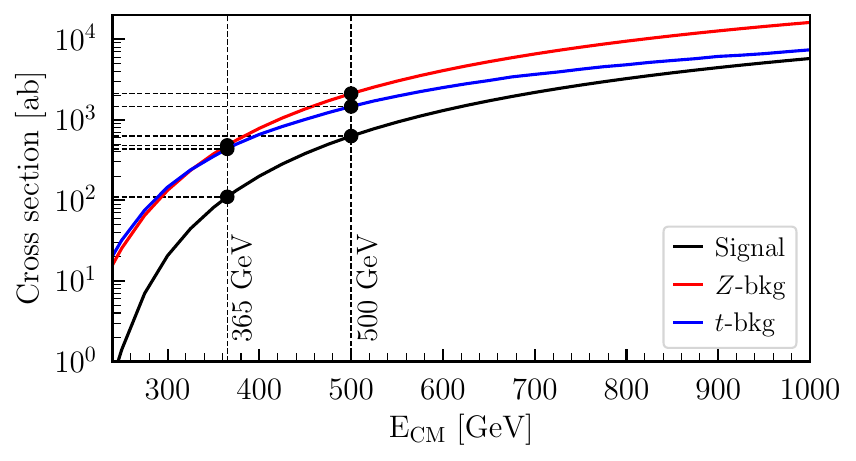}
\end{center}
\caption{Variation of total cross section as a function of center-of-mass energy of the FCC-ee.}
\label{fig:crossx-ecm}
\end{figure}

As mentioned, the FCC-ee is proposed to operate at several center-of-mass energy points, {\it viz.} the $Z$-pole, $WW$ threshold, $ZH$ threshold, and $t\bar t$ threshold. In this work, we mainly focus on the $t\bar t$ threshold corresponding to a center-of-mass energy of 365~GeV with a projected luminosity of 0.67~ab$^{-1}$ per year. This operating point is particularly suitable, as it provides a sufficiently large event sample to study the signal process.

We first examine the cross sections of the signal and the two major background processes as a function of the center-of-mass energy in Fig.~\ref{fig:crossx-ecm}. The corresponding cross sections at $\sqrt{s}=365$~GeV are then listed in Tab.~\ref{tab:crossx}. In Fig.~\ref{fig:crossx-ecm} and Tab. ~\ref{tab:crossx}, the signal and \Zbkg\ have been generated without any cut, whereas for \tbkg\ a minimum of $p_T=5$ GeV cut has been implemented to avoid a t-channel singularity. The cross-section for signal is $111.3 \:\text{ab}$, and for \Zbkg\ and \tbkg\ are $480.0\:\text{ab}$ and  $422.8\:\text{ab}$ respectively for $365$ GeV $E_{\rm CM}$.  The scale uncertainty with $m_H/2\leq\mu_0(m_H)\leq2m_H$ for the signal is $\sim\pm 6\%$. The scale uncertainties for the backgrounds are comparable to those of the signal. For $500$ GeV $E_{\rm CM}$, the signal cross-section is $634.5$ ab, whereas the background cross-sections are $2100.7$ ab and $1416.0$ ab for \Zbkg\ and \tbkg\ respectively. Although background processes have slightly larger cross sections, the characteristic peak of the invariant mass of $b\bar b$ near the Higgs boson mass would help to separate the signal from the other backgrounds. 

\begin{table}
\begin{center}
\begin{tabular}{|c||r|r|r|}
\hline
\bf \multirow{2}{*}{$E_{\rm CM}$} & \multicolumn{3}{|c|}{Cross section [ab]}\\
\cline{2-4}             & Signal & \Zbkg & \tbkg \\
\hline
\bf 365~GeV & $111.3$ &  $480.0$ &  $422.8$\\
    500~GeV & $634.5$ & $2100.7$ & $1416.0$\\
\hline
\end{tabular}
\end{center}
\caption{Cross section for signal and background process at $365$ GeV FCC-ee. This includes the branching ratios of $H\rightarrow b \bar b$ for signal, $Z \rightarrow b\bar b$ and $t (\bar{t}) \rightarrow W^+ b (W^-\bar b)$ for backgrounds, respectively.} 
\label{tab:crossx}
\end{table}

The generated parton-level hard interaction events stored in LHE format are then passed through {\tt Pythia8}\,\cite{Bierlich:2022pfr} for QCD and electroweak showering, followed by hadronization of colored partons. We subsequently perform a fast detector simulation using {\tt Delphes}\,\cite{deFavereau:2013fsa} with detector modeling implemented through the FCC-ee Delphes card\,\cite{delphes_FCCee_card_git}. Jets are clustered from detector-level hadrons using {\tt FastJet3}\,\cite{Cacciari:2011ma} with the anti-$k_t$ algorithm\,\cite{Cacciari:2008gp} and a radius parameter of 0.6.
We focus on the Higgs boson in its most favorable decay mode in $b \bar b$, and therefore expect to have two $b$-tagged jets in the final state. Tagging a $b$-jet requires sophisticated techniques and detailed detector specifications, which are currently not readily available for the FCC-ee.
To account for this, we assume a flat 85\% tagging efficiency for $b$-jets originating from $b$-quarks and a mistagging rate of 40\% for $c$-quarks misidentified as $b$-jets. These estimates are based on current results from the CMS collaboration\,\cite{CMS:2017wtu,Mitra:2022qcu,Pfeffer:2023qet} at the LHC Run 2 and are expected to improve in the cleaner FCC-ee environment, along with more advanced techniques. The chance of mistagging light quarks or gluons as $b$-jets is negligible and is therefore not included in the background estimation.
\begin{figure}[h]
\begin{center}
\includegraphics[width=\columnwidth]{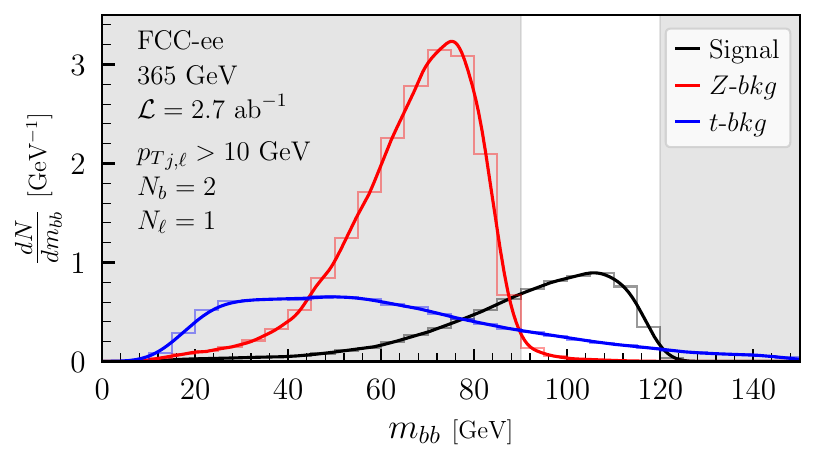}
\end{center}
\caption{Distribution of two $b$-jets invariant mass $m_{bb}$.}
\label{fig:mbb}
\end{figure}

In the analysis, we first identify exactly one charged lepton ($\ell^\pm \in \{ e^\pm, \mu^\pm$\}) with $p_T \ge 10$~GeV and exactly two $b$-tagged jets with $p_T\ge 10$~GeV. These $p_T$ thresholds are relatively lower than those typically used at the LHC, but are well motivated by the clean environment of lepton colliders, with significantly reduced contamination from underlying events and pile-up. Hence, a 10 GeV lower fiducial cut is reasonable. With these event selections, we first show in Fig.~\ref{fig:mbb} the invariant mass distribution of the two $b$-jets for the signal and the two dominant backgrounds, {\it viz.} \Zbkg and \tbkg. Each distribution is normalized to the expected number of events at $\mathcal{L}=2.7$ ab$^{-1}$, such that the area under each curve corresponds to the total event yield. The solid curves are obtained using Kernel Density Estimation (KDE) with Gaussian kernels and Scott's bandwidth\,\cite{scott_multivariate_1992}. The corresponding histograms are overlaid with faded lines, ensuring consistency between the two representations.

From Fig.~\ref{fig:mbb}, we can see that the two peaks corresponding to $H$ and $Z$ masses are well separated, while the background \tbkg shows a continuously falling spectrum.
One also notices that the two resonant peaks corresponding to the Higgs and $Z$ boson masses are shifted towards lower values. This is primarily due to the detector measurement error, which typically measures lower energy compared to the true values.
In actual experiments, advanced techniques such as jet energy scale or jet energy resolution, as implemented in the CMS\,\cite{CMS:2016lmd,CMS:2017wyc} or ATLAS\,\cite{ATLAS:2017bje,ATLAS:2020cli,ATLAS:2023tyv,ATLAS:2024kkj} experiments at the LHC, are used to correct for detector effects.
These corrections can restore the peak positions closer to the true Higgs or $Z$ boson masses. In this analysis, however, we adopt a conservative approach and work with uncorrected jet energies, anticipating that the inclusion of such techniques would only improve the results.
\begin{figure}[h]
\centering
\includegraphics[width=\columnwidth]{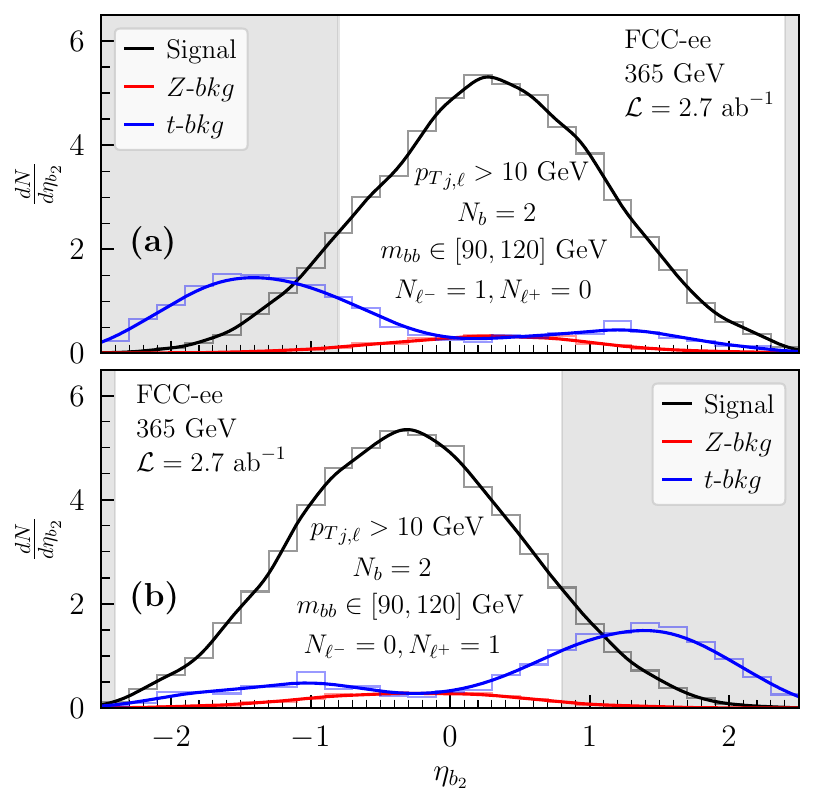}
\caption{Distribution of $\eta$ of subleading $b$-jet.}
\label{fig:etab2}
\end{figure}

A selection criterion of $m_{bb} \in [90,120]$~GeV, as shown by the unshaded region in Fig.~\ref{fig:mbb}, significantly reduces the \Zbkg background, although some \tbkg background still remains. To further suppress this contribution, we consider another discriminating variable, {\it viz.} $\eta_{b_2}$, the pseudorapidity of the subleading $b$-jet. In the \tbkg background, the subleading $b$-jet often arises directly and not from the top decay, and is therefore expected to be more forward, leading to larger pseudorapidity.

This is seen in Fig.~\ref{fig:etab2}(a), where the distribution of $\eta_{b_2}$ is shown for events in which an $e^-$ or $\mu^-$ (negatively charged lepton) is identified in the final state. One observes that the $\eta_{b_2}$ distribution is asymmetric around zero for events with a negatively charged lepton. This asymmetry is expected due to the asymmetry in the PDFs of the electron and the photon originating from the electron and positron beams, respectively. A similar behavior is observed for events with a positively charged lepton in the final state, as shown in Fig.~\ref{fig:etab2}(b).
Therefore, we impose an asymmetric cut on $\eta_{b_2}$.
To apply this, one needs to identify the sign of the charged lepton. The optimized cut we find is $\eta_{b_2} \in [\pm0.8,\mp2.4]$ for the events having $\ell^\pm$ in the final state, and they are depicted by unshaded regions in Fig.~\ref{fig:etab2}.

\begin{figure}[h]
\centering
\includegraphics[width=\columnwidth]{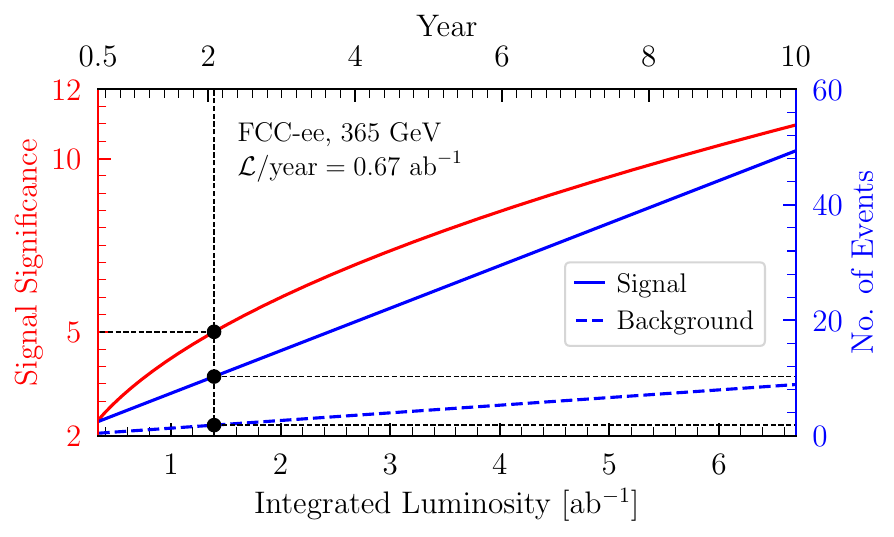}
\caption{Significance as a function of integrated luminosity (bottom x-axis) and as a function of year (top x-axis) for $E_{\rm CM}=$ 365~GeV FCC-ee. The number of signal (solid) and background (dashed) events is also plotted in blue and can be read from the right y-axis. A 5$\sigma$ significance is achievable in 2 years. }
\label{fig:signi-lumi}
\end{figure}

We now move on to estimate the signal significance. For a given number of signal events $S$ and background events $B$, the signal significance is calculated using the most general formula~\cite{Cowan:2010js}:
\begin{equation}
\mathfrak{S} = \sqrt{2(S+B)\ln(1+S/B)-2S}.
\end{equation}
We use this expression since both $S$ and $B$ are small and do not satisfy the $S/B \ll 1$ condition, for which the commonly used approximation $S/\sqrt{B}$ is applied. The signal significance is evaluated assuming $\mathcal{L}/{\rm year}=0.67$ ab$^{-1}$, as projected by the FCC-ee study group\,\cite{FCC:2025uan}. In Fig.~\ref{fig:signi-lumi}, we first show the signal significance as a function of integrated luminosity (red solid curve) for 365~GeV FCC-ee. The blue solid and dashed curves denote the number of signal and background events, respectively, as indicated on the right $y$-axis. The top axis shows the corresponding number of years required to reach a given luminosity. The solid black point on the red curve corresponds to a 5$\sigma$ significance, which can be achieved in about 2 years of running at the 365~GeV.

\begin{figure}[h]
\centering
\includegraphics[width=\columnwidth]{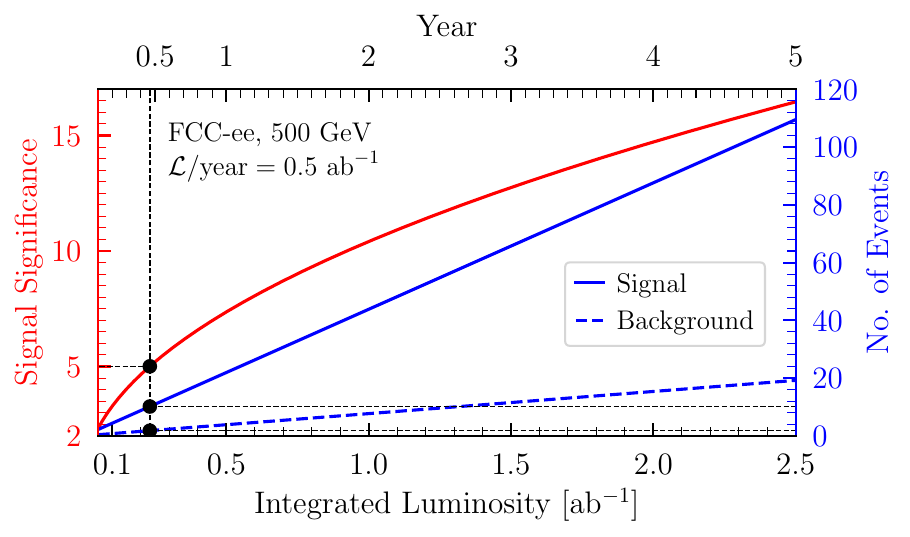}
\caption{Significance as a function of integrated luminosity (bottom x-axis) and as a function of year (top x-axis) for $E_{\rm CM}=$ 500~GeV FCC-ee. The number of signal (solid) and background (dashed) events is also plotted in blue and can be read from the right y-axis. A 5$\sigma$ significance is achievable within a few months.}
\label{fig:signi-lumi_500}
\end{figure}

We then analyze the same signal and background processes but for $E_{\rm CM} = 500$~GeV, for which the $m_{bb}$ and $\eta_{b_2}$ distributions are explicitly not shown. We find that the same set of cuts optimizes the signal significance at this energy as well. The cuts are ${p_T}_{j,\ell} > 10$~GeV, $N_b=2$, $N_{\ell^\pm} =1$, $m_{bb} \in [90,120]$~GeV, and $\eta_{b_2} \in [\pm0.8,\mp2.4]$. The variations of signal significance, along with the number of signal and background events as a function of integrated luminosity, is shown in Fig.~\ref{fig:signi-lumi_500}. Assuming a yearly luminosity of $0.5$~ab$^{-1}$, the corresponding running year is also indicated on the top x-axis. With these considerations, this machine can achieve a 5$\sigma$ discovery of the process within a few months of running. 

While there is scope for further improvement once the detector design is finalized and advanced analysis techniques are incorporated, our simplistic analysis already demonstrates that this novel Higgs production channel, associated with a lepton from the $W$-boson, can be discovered at 5$\sigma$ significance in about 2 years at 365~GeV and within a few months at 500~GeV, assuming luminosities of 0.67 and 0.50~ab$^{-1}$ per year, respectively. While we demonstrate one example of a lepton-photon initiated process in this work, it opens up the possibility of a broader class of processes with similar initial states, which can contribute significantly both to signal and background at higher energies. This strongly motivates the inclusion of such processes in the physics program planning, design of experiments, trigger strategies, and analysis frameworks.

In this letter, we have considered a novel production mechanism for the Higgs boson. After the HL-LHC, FCC-ee is most 
likely to operate in the near future at the center-of-mass energy of about $365$ GeV. At this machine, a new mechanism to produce the 
Higgs boson is $e^\pm \gamma \rightarrow  {\bar \nu}_e (\nu_e) H W^\pm$. We have shown that the signature of this process, `isolated charged lepton and two $b$-jets', is relatively background-free.  We have used an in-house MC event generator to deal with the technical
challenge of simultaneously incorporating the photon PDF and electron/positron PDF,
which has singular behavior near the high momentum fraction region.
We find that the channel can be observed at 5$\sigma$ significance over the other Standard Model backgrounds within two years of operation at the FCC-ee 365 GeV mode with a yearly integrated luminosity of 0.67 ab$^{-1}$. However, with better reconstruction
of bottom jets and better detector efficiencies, one can reduce the time for the observation of the signal process and studying the
Higgs boson properties. We also show that a machine operating at the higher center-of-mass energy, say, $500$ GeV, will also help to probe this process, and the $5\sigma$ level significance will be reached within a few months.

{\it Acknowledgments:} B.~D.~and T.~S.~acknowledge Sven-Olaf Moch and V.~Ravindran for the useful discussions.
T.~S. also acknowledges useful discussions with Soureek Mitra and Mrinal Kanti Pal. 
The authors acknowledge the HPC facility Kamet at The Institute of Mathematical Sciences for various computational needs.
T.~S.~acknowledges Vishal Bhardwaj and Ambresh Shivaji for their hospitality during his visit to IISER Mohali.


\providecommand{\href}[2]{#2}\begingroup\raggedright\endgroup

\end{document}